# Biometric recognition: why not massively adopted yet?[1]

Marcos Faundez-Zanuy
Escola Universitaria Politècnica de Mataró
Avda. Puig i Cadafalch 101-111
08303 MATARO (BARCELONA) SPAIN
E-mail: faundez@eupmt.es http:www.eupmt.es/veu

**ABSTRACT**
Although there has been a dramatically reduction on the prices of capturing devices and an increase on computing power in the last decade, it seems that biometric systems are still far from massive adoption for civilian applications. This paper deals with the causes of this phenomenon, as well as some misconceptions regarding biometric identification.

**INTRODUCTION**

Biometric systems offer some advantages over the classical handheld tokens (card, ID, passport, etc.) and knowledge-based (password, PIN, etc.) authentication methods. However, it seems that classical systems are not being replaced by biometric ones. According to [1], this is due to four fundamental problems:
1) Security: How to guarantee that the sensed measurements are not fraudulent.
2) Privacy: How to make sure that the application is indeed exclusively using pattern recognition for the expressed purpose.
3) Accuracy: How to accurately and efficiently represent and recognize biometric patterns.
4) Scale: How to acquire repeatable and distinctive patterns from a broad population.

This paper intends to discuss these four aspects providing new points of view. In addition, we will try to beat some misconceptions regarding biometric recognition, using the experience and knowledge of some other fields, such as anthropology and law.

The problem concerning security was described in detail in [2]. One of the main critical facts is that when biometric identifiers have been compromised, the legitimate user has no mean to revoke the identifier in order to switch to another new one. This is a serious drawback when compared with a handheld token or a knowledge based verification method, which can be replaced. Probably one solution should be a change on the philosophy: the biometric pattern can be considered as a "login" instead of a "password", or a login plus biometric plus password based system. Thus, the recognition relies on both information: login and password, where the latest can be kept secret and changed if necessary. This lets to improve the security, although loses the nice property of biometric authentication systems, where it is not necessary to remember anything neither to hold any card. However, the engineers still try to use the biometric signal as password. Another solution is the combination of several biometric systems [3]. In this case it is more difficult for a hacker to fake several systems than a single one. If there is a human supervisor during the acquisition of the biometric signal, the possibility to introduce a fraudulent signal is almost negligible. An example of this situation would be border entrance control.

The problem concerning privacy was described in detail in [4]. Probably the best way to preserve privacy is through regulations and fines to those who break it. Thus, it is the work of legislators to deal with it. This is not a problem specific to biometrics, and strong efforts are dedicated to set up personal data protection laws.

The last two problems are mainly related to algorithmic issues and point out that biometric identification is not a solved problem yet. Thus, more research must be done in order to improve current systems. However, is biometric recognition as easy as we think?

The remaining part of this paper gives some arguments about why performing biometric recognition is not as easy as we think. For this purpose we will use some arguments taken from other fields, probably more matured than biometrics.

---

[1] This work has been supported by FEDER and MCYT, TIC-2003-08382-C05-02

**BIOMETRIC RECOGNITION: AN EASY TASK?**

Before having an automatic system able to solve our identification problems with the accuracy and scale that we desire, we can examine the situation when the operator is a human being. Probably if a person can perform this task, and we can summarize the procedure done by him/her, the answer is yes. But what does really mean that "a human can do the task easily"? In this section we describe some facts that can help understand the complexity of biometric recognition.

**Face Recognition: an easy task for human beings?**

Although common sense points out that face recognition is an easy-task for human beings, it is more complex than we think. It is not an innate ability, at least when using two-dimensional representation of faces (photos, drawings, etc.), and it must be acquired when we are young. A British anthropologist, named Nigel Barley, went to study the Dowayo people (see figure 1), a strangely neglected group in North Cameroon, and he reports in his book [5]:

*"…In the end I managed to lay my hands on some postcards depicting African fauna. I had at least a lion and a leopard and showed them to people to see if they could spot the difference. Alas, they could not. The reason lay not in their classification of animals but rather in the fact that they could not identify photographs. It is a fact that we tend to forget in the West that people have to learn to be able to see photographs. We are exposed to them from birth so that, for us, there is no difficulty in identifying faces or objects from all sorts of angles, in differing light and even with distorting lenses. Dowayos have no such tradition of visual art; theirs is limited to bands of geometric designs. Nowadays, of course, Dowayo children experience images through schoolbooks or identity cards; by law, all Dowayos must carry an identity card with their photograph on it. This was always a mystery for me since many who have identity cards have never been to the city, and there is no photographer in Poli. Inspection of the cards shows that often pictures of one Dowayo served for several different people. Presumably the officials are not much better at recognizing photographs than Dowayos…*

*…The point was that men could not tell the difference between the male and the female outlines. I put this down simply to my bad drawing, until I tried using photographs of lions and leopards. Old men would stare at the cards, which were perfectly clear, turn them in all manner of directions, and then they say something like "I don't know this man". Children could identify the animals…."*

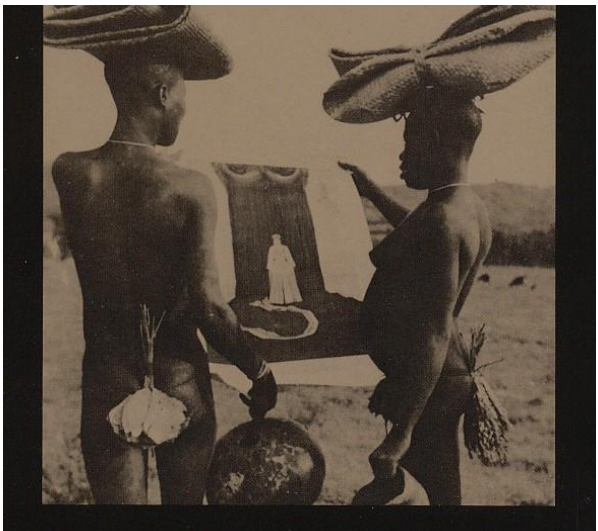

**Fig. 1. Dowayos looking at drawings**

Thus, the face recognition ability of human beings is not as trivial as we could think. It is the result of a long learning process, which is not really well understood, varies along our life, and even relies on different parameters [6]. Probably it is similar to learn to pronounce some phonemes. It is not a problem when we are child, but after several years, it is almost impossible (for instance, a French person cannot pronounce the /rr/ phoneme of the Spanish word "perro",

___

which means dog). Taking into account the experience of this anthropologist, we can see that really the enrollment phase, where the computer tries to learn a model for each person using some (limited amount of) training data, is fundamental.

Even when we are used to seeing photos and recognizing faces, there are difficult situations. The first coming to our mind is that related to changes in hairstyle, makeup, facial hair, addition or removal of eyeglasses, hats, scarves, etc. Figure 2 shows six different photographs of the actor Val Kilmer on the film "The saint". Although it is an extreme case, where the user is trying to avoid his identification, most faces databases contemplate changes on facial expression, illumination, etc. because it is an actual problem in real scenarios.

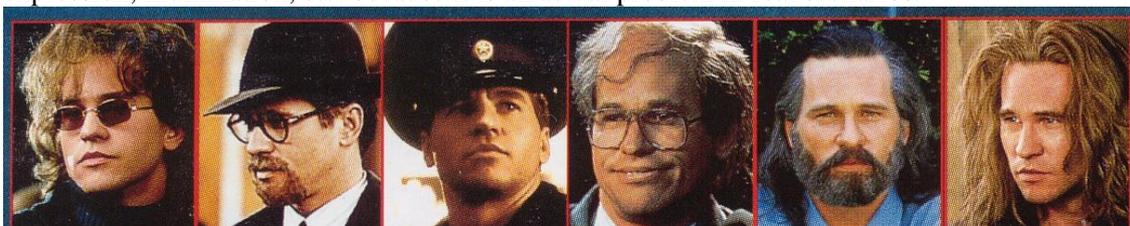

**Fig. 2 Six different snapshots of Val Kilmer.**

Another difficult situation is when trying to recognize a human face different from the race that we are used to deal with in our normal life. For instance, [7] reports that "*The last half-century's empirical studies of cross racial IDs has shown that eyewitnesses have difficulty identifying members of another race...*"

On the other hand, human beings are also subject to errors on our face identification decisions, even more than we think! This is well-known on justice courts. One of the most famous cases [8] of mistaken identity occurred in the 1896 English trial of Adolf Beck. Beck was convicted based on the identifications of ten women. He was convicted and spent 7 years in prison, all the while maintaining his innocence. He claimed he was mistaken for a man named John Smith. While Beck was still in prison, more of these offenses occurred. Smith was eventually arrested and Beck was released. Afterwards it was discovered that Beck had spent seven years in prison for a crime he did not commit, a committee was formed. The committee found that "*evidence as to identity based on personal impressions, however bona fide, is perhaps of all classes of evidence the least to be relied upon, and therefore, unless supported by other facts, an unsafe basis for the verdict of a jury.*"

Thus, [8] concludes that "*No person should be deprived of his liberty solely on the basis of eyewitness testimony unless the jury is fully aware of the ways in which such testimony may be flawed*".

The conclusion of this section should be that things are not as easy as they seem, and for some recognition systems, we cannot expect 100% identification rates neither 0% false acceptance and rejection rates.

**Are human beings better than machines for biometric identification?**

We should not be surprised if automatic face recognition systems exhibit some amount of errors. This is not exclusive of face recognition technology. In [9-10] we found that human beings outperform computers for the task of locating the characteristic points of a fingerprint. Thus, there is still room for improvement on this technology, and even better results can be expected. However, in fingerprints it is easy to define what we are looking for. Mainly it is terminations and bifurcations on the ridges of the fingerprint [11-12]. For other biometric patterns, a human being can experiment troubles in identifying people, because they are not used as identifiers in our normal life. For instance, figure 3 shows four snapshots of four hands. How many different people can we differentiate? If, instead of four hand scanned images, there were four photos of their face, the answer would be trivial.

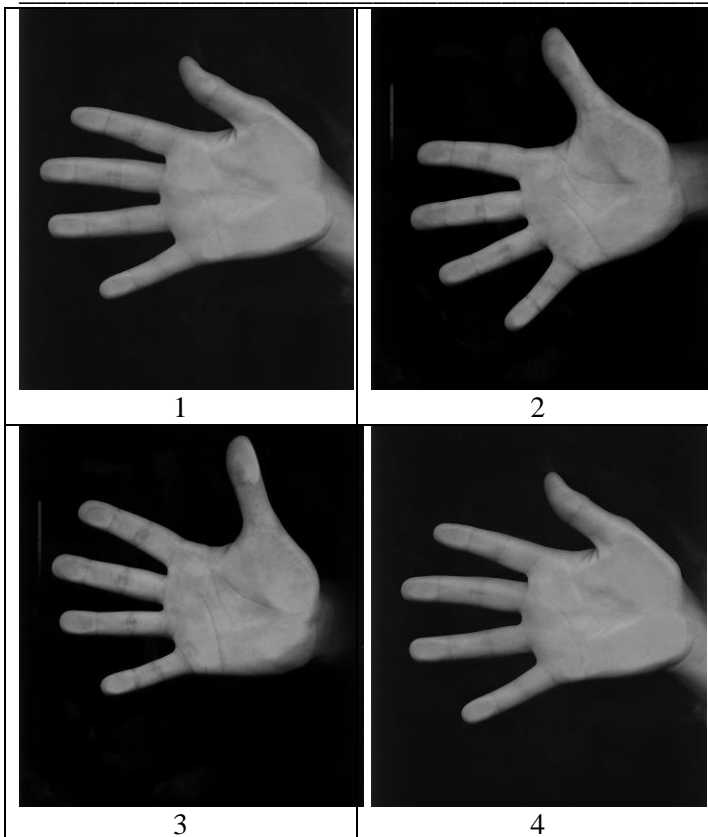

**Fig. 3. Four different snapshots of a hand. How many different persons are?**

For this quiz, plots number 1 and 4 really belong to the same female person, and 2 and 3 to another different one. In this case, it is easier for a computer to work out some parameters like finger length, width, etc. and assign identities, than for a human being. Thus, there certainly are situations where a computer can outperform a person. In any case, the adoption of a biometric solution can dissuade criminals, and it is nowadays good enough for a huge set of civilian applications. Although it is not a trivial problem, let us hope much better solutions in a near future.